\documentstyle[12pt]{article}
\begin{document}

{\Large\bf{\centerline{On the constrained structure of duality symmetric
Maxwell theory}}}
\bigskip
\centerline{R. Banerjee{\footnote{e-mail: rabin@boson.bose.res.in}}}
\bigskip
\centerline{S.N. Bose National Centre for Basic Sciences}
\centerline{Block- JD, Sector III, Salt Lake City}
\centerline{Calcutta- 700,091; India}
\bigskip
\begin{abstract}
The constrained structure of the duality invariant form of
Maxwell theory is considered in the Hamiltonian formulation of
Dirac as well as from the symplectic viewpoint. Compared to the
former the latter approach is found to be more economical and
elegant. Distinctions from the constrained analysis of the usual
Maxwell theory are pointed out and their implications are also
discussed. 
\end{abstract}
\newpage
\section{\bf Introduction}
\bigskip
The relevance of duality symmetry in either field or string
theories has been reviewed \cite{R} in various contexts. The duality
invariant Lagrangeans obtained in different cases, though
equivalent to their parent Lagrangeans, present features which
are distinctly unique. While most investigations \cite{R, DT,
SS, BW1} are directed
towards undersatnding the nature of duality symmetry in such
Lagrangeans, a detailed constrained analysis is lacking. Some
sporadic computations \cite {G} are available but these are incidental and
are merely used for getting other results.

In this paper a detailed constrained analysis of the duality
invariant form of Maxwell theory is done both in the Hamiltonian
formalism of Dirac and in the symplectic approach, which is a
Lagrangean formulation. We find that the constrained structure
is quite distinct from the conventional Maxwell theory.
Moreover, the results are obtained in a more clean and brief
fashion in the symplectic approach as compared with the Dirac approach.

In section 2, the Dirac analysis is performed. The theory
presents a  mixed system containing both first and second class
constraints. The constraints in the second class set are not
independent thereby providing an example of a reducible system.
By introducing auxiliary variables \cite{BB} this reducibility is taken
into account and the second class constraints are eliminated by
computing the relevant Dirac brackets. Interestingly, it is
observed that the knowledge of these brackets is sufficient to
obtain the complete algebra among the gauge invariant variables
and explicit reference to the first class constraints or any
gauge fixing is avoided. The results are found to agree with the
standard analysis \cite{G} where the complete constraint sector
including the Coulomb gauge was taken.
We also show how to implement
the axial gauge condition in the present context. The
independent canonical pairs are easily identifiable in this gauge.

The symplectic treatment of the model is carried out in section
3. Once again there are striking differences from the usual
constraint analysis of the Maxwell theory. It is found that the
zero modes of the non-invertible symplectic two form do not lead
to any new condition. This signals the presence of a gauge theory.
However, none of the constraints- either first or second class-
found in the Dirac scheme are obtained here. Effectively,
therefore, the model simulates the features of an unconstrained
theory. The noninvertibility of the symplectic two form is due
to the fact that the duality invariant Lagrangean is expressed
solely in terms of the transverse components of the gauge
potentials, quite in contrast to the usual Maxwell Lagrangean.
Incorporating this transversality in the Lagrangean, either by means of a
Lagrange multiplier or directly, the symplectic two form becomes
invertible. The brackets are then easily read-off from the
inverse. Since constraints do not emerge at any stage of the
computations, the issue of gauge fixing is completely bypassed.
Nevertheless, we explicitly show how the different gauge fixed
results in the Dirac approach are reproduced in the symplectic
formalism. 

Some concluding remarks have been made in section 4.

\bigskip

\section{\bf Dirac Analysis}
\vskip 0.5cm
In this section a detailed Dirac analysis of the duality
symmetric electromagnetic Lagrangean will be performed. In terms
of the pair of electric and magnetic fields this Lagrangean is given by
the familiar expression \cite{SS, BW1, G},
\begin{equation}
{\cal L} = \frac{1}{2}( B^a_i\epsilon_{ab} E^b_i
 - B^a_i B^a_i)
\label{d1}\end{equation}
where,
\begin{equation}
{E_i}=-F_{0i} = -\partial_0A_i + \partial_iA_0
\label{d2}
\end{equation}
\begin{equation}
{B_i}=\epsilon_{ijk}\partial_jA_k
\label{d3}\end{equation}
and $a=1, 2$ is an internal index charactersing the pair of fields while
$\epsilon_{12}=-\epsilon_{21}=1$. Expressed in terms of the potentials the
Lagrangean, modulo a divergence term, simplifies to,
\begin{equation}
{\cal L} ={\frac{1} {2}}\epsilon^{jki}\partial_j{A_k}^{a}\epsilon_{ab}
\partial_{0}{{A^b}_{i}}- \frac{1}{4} F^{a,jk} F^a_{jk}
\label{d4} \end{equation}

The above Lagrangean will be the starting point of the Dirac
constrained analysis. Note that the time component of the
potentials are absent so that we may, without any loss of
generality, do away completely with the canonical set $(A_0,
\pi^0)$. In that case the primary constraints of the system are
given by,
\begin{equation}
\Omega^a_{i} = \pi^a_{i} + 
\frac{1}{2}\epsilon_{a b}
\epsilon_{ijk}\partial_jA^b_{k}\approx 0
\label{d7}
\end{equation}
It can be checked that no further constraints are generated by
demanding the time conservation of (\ref{d7}). Furthermore, the
above set is a mixture of first and second class constraints.
These may be easily separated by decomposing (\ref{d7}) into its
transverse and longitudinal components \cite{G}. Then the first class
constraints are given by,
\begin{equation}
\partial_i\Omega_i^a=\partial_i\pi_i^a\approx 0
\label{d8}
\end{equation}
while the second class ones are,
\begin{equation}
\Omega^a_{T, i} = \pi^a_{T, i} + 
\frac{1}{2}\epsilon_{ab}
\epsilon_{ijk}\partial_jA^b_{T, k}\approx 0
\label{d9}
\end{equation}
In order to compute the Dirac brackets the usual approach is to
fix a gauge corresponding to the first class constraints
(\ref{d8}). The Coulomb gauge, 
\begin{equation}
\partial_iA_i^a\approx 0
\label{d10}
\end{equation}
is a popular choice. Now the complete constraint sector is given
by (\ref{d8}), (\ref{d9}) and (\ref{d10}). Since the constraint
structure is rather involved, an elaborate
calculation is now necessary to explicitly get the Dirac
brackets.{\footnote {These will be denoted by a star}}
The results, which are given in terms of the transverse
components since these are the gauge invariant variables, 
are quoted from the literature \cite{G},
\begin{equation}
\{A^a_{T,i}(x) ,A^b_{T,j}(y)\}^*  = \epsilon_{ab}\epsilon_{ijk}
\frac{\partial_k}{\bf \nabla^2} \delta(x-y)
\label{d11}\end{equation}
\begin{equation}
\{A^a_{T,i}(x) ,\pi^b_{T,j}(y)\}^*  =\frac{1}{2}\delta_{ab} 
\Big(g_{ij}+\frac{\partial_i\partial_j}{\bf \nabla^2}\Big) \delta(x-y)
\label{d12}\end{equation}
\begin{equation}
\{\pi^a_{T,i}(x) ,\pi^b_{T,j}(y)\}^*  =- \frac{1}{4}
\epsilon_{ab}\epsilon_{ijk}
\partial_k \delta(x-y)
\label{d13}\end{equation}
where,
\begin{equation}
{A^{a}_{T,i}}{(x)} = ({\delta_{ij}}-\frac{\partial_i \partial_j}
{\nabla^2}){A^{a}_{j}}{(x)}
\label{s8}\end{equation}
All the constraints are now strongly implemented.

We now discuss an alternative formulation of the problem which
highlights certain distinctive features that are otherwise
hidden. Let us begin with the second class set (\ref{d9}). 
Confined to this set, the Poisson bracket matrix is given by,
\begin{equation}
\{\Omega^{a}_{T,i}(x) ,\Omega^{b}_{T,j}(y)\}_{PB} = 
\epsilon_{ab} \epsilon_{ijk} \partial_k \delta(x-y)
\label{d14}\end{equation}
The right hand side has no inverse which, at first sight, would
seem bit surprising since we are considering only the second
class sector. The reason for this anomalous behaviour is that
the second class constraints are reducible because,
\begin{equation}
\partial_i\Omega^a_{T, i}=0
\label{d15}
\end{equation}
so that all $\Omega^a_{T, i}$ are not independent. The
standard way is to isolate the independent subset of constraints
and proceed with the computations. Besides being messy, such an
approach is generally not recommended since certain symmetries
of the original problem may be lost. A
particularly elegant way is to suitably modify the constraints
by introducing an additional pair of canonical variables \cite{BB}, 
\begin{equation}
{\tilde \Omega}^{a}_{T,i}(x) =\pi^{a}_{T,i} + {\frac{1}{2}}\epsilon_{a b}
\epsilon_{ijk}\partial_jA^b_{k,T} + \partial_i \phi^{a}
\approx 0
\label{d16}
\end{equation}
where, 
\begin{equation}
\{\phi^{a}(x) ,\phi^{b}(y)\} = \frac{\epsilon _{ab}}{m} \delta(x-y)
\label{d17}
\end{equation}
and a mass scale has been introduced for dimensional reasons.
It is easy to see that $\partial_i{\tilde
\Omega}^{a}_{T,i}\approx 0$ leads to
$\phi^a\approx
0$ so that the modified constraint is weakly equivalent to the
original one. Moreover, by this modification, the reducibility
in the initial constraint has been removed. We can now directly
work with the new constraint. A similar strategy was used
earlier to analyse in details the reducibility in the
constrained structure of $p$- form gauge fields \cite {BB}.
The  Poisson bracket matrix among the new constraints is
now given by,
\begin{equation}
{{\large C}^{ab},}_{ij}=
\{{\tilde \Omega^{a}}_{T,i}(x) ,{\tilde \Omega^{b}}_{T,j}(y) \} =
 \epsilon_{ab}[\epsilon_{ijk} \partial_k - {\frac{1}{m}} 
\partial_i\partial_j] \delta(x-y)
\label{d18}
3\end{equation}
which has the following inverse,
\begin{equation}
 { {\large C}^{ab},}_{ij}^{-1} =
 -\epsilon_{ijk} {\frac{\partial_k}{\nabla^2}}
- m {\frac{\partial_i \partial_j}{(\nabla^2)^2}}
\label{d19}
\end{equation}
It is now straightforward to calculate the Dirac brackets among
the  basic variables. The result is,
\begin{equation}
\{A^a_i(x) ,A^b_j(y)\}^*  = \epsilon_{ab}\epsilon_{ijk}
\frac{\partial_k}{\bf \nabla^2} \delta(x-y)
\label{d11a}\end{equation}
It is simple to prove that the complete algebra (\ref{d11},
\ref{d12}, \ref{d13}) is reproduced from the above equation and
using the strong implementation  of the second class constraints.

It is worthwhile to note that  by properly accounting for the second class
sector only, the complete Dirac algebra among the physical
variables has been obtained. Explicit reference to either the
first class constraint or any gauge fixing has been avoided.

In usual electrodynamics the axial gauge is sometimes used since
it clearly exposes the canonical pairs of the theory \cite{HRT}. Let us see
these features in the present context. The full constraint
sector has to be considered. Apart from the second class
constraints and the Gauss constraint, the additional constraint
is given by,
\begin{equation}
A_3^a\approx 0
\label{d22}
\end{equation}
The matrix of the Poisson brackets among the constraints
$\Omega_i^a$ is now
given by,
\begin{equation}
\{\Omega_i^a(x), \Omega_j^b(y)\}=
{\bf \large C}^{ab} = \pmatrix{ 0& {\delta_{ab} \partial_3} 
& 0 & 0
 \cr {\delta_{ab} \partial_3} & 0 
& {\delta_{ab} \frac{\partial_1 \partial_3}
{\nabla^2}} &  {\delta_{ab} \frac{\partial_2 
\partial_3}{\nabla^2}}  
\cr 0&  - {\delta_{ab} \frac{\partial_1 \partial_3}{\nabla^2}} & 0 
& \epsilon_{ab}\partial_3  \cr
 0  & -{\delta_{ab} \frac{\partial_2 \partial_3}{\nabla^2}} 
&
- \epsilon_{ab} \partial_3 & 0 } \times \delta(x-y)\label{d23}
\label{n1}
\end{equation}
where the first two constraints $(i=1, 2)$ refer to the Gauss law and the
axial gauge, respectively, while the last two $(i=3, 4)$ are the 
second class set. The inverse matrix is given by,
\begin{equation}
{\bf \large C^{-1}} =
\pmatrix { 0 & \delta_{bc}\frac{1}{\partial_3}  &
  \epsilon_{bc} \frac{\partial_2}{\nabla^2 \partial_3}
 &  - \epsilon_{bc} \frac{\partial_1}{\nabla^2 \partial_3}
\cr \delta_{bc} \frac{1}{\partial_3} & 0 & 0 &0 
\cr \epsilon_{bc} \frac{\partial_2}{\nabla^2 \partial_3}
 & 0 & 0 & \frac{1}{\partial_3} \epsilon_{bc}
\cr - \epsilon_{bc} \frac{\partial_1}{\nabla^2 \partial_3}
& 0 & - \frac{1}{\partial_3} \epsilon_{bc} & 0 }
 \times \delta(y-z)\label{d24}
\label{n2}
\end{equation}
The Dirac brackets in the axial gauge are now easily found,
$$
 \{{ A^{a}_{1}}(x) , {A^{b}_{2}}(y) \}^* =
\{{ A^{a}_{1}}(x) , {A^{b}_{2}}(y) \} -
 \int \{{ A^{a}_{1}}(x) ,{{\Omega^c}_i}(z)\}
{{\bf \large C}^{cd}_{ij}}^{-1}
\{{{\Omega^d}_j}(w) ,{A^{b}_{2}}(y) \}
$$
\begin{equation}
= \epsilon_{ab} \frac{1}{\partial_3} \delta(x-y)\label{d25}
\label{n3}
\end{equation}
Two observations are now in order. First, the brackets among the
transverse variables,
\begin{equation}
\{{ A^{a}_{T,1}}(x) ,{ A^{b}_{T,2}}(y)\}^*
= \epsilon_{ab} \frac{\partial_3}{\nabla^2} \delta(x-y)\label{d26}
\label{n4}\end{equation}
reproduce the Coulomb gauge result (\ref{d11}). This acts as a
consistency check since the above algebra must be gauge
independent. Secondly, it is simple to identify the canonical
pairs as $A_1, \pi^1$ and $A_2, \pi^2$ since, using the
definition for the momenta and (\ref{d25}), it follows that,
\begin{equation}
\{A^{b}_{1}(x) , \pi^{1}_{c}(y)\}^* 
=  \{ {A^{b}_{2}}(x) , {\pi^{2}_{c}}(y)\}^*=
\frac{1}{2}\delta_{bc}
\delta (x-y)\label{d27} 
\label{n5}
\end{equation}
The analogous situation in usual Maxwell theory may be recalled
where these set of variables (without the internal indices)
characterise the canonical pairs \cite{HRT}.

\bigskip

\section{\bf Symplectic Analysis }

\bigskip

The duality symmetric Maxwell theory is now analysed in the
context of the  symplectic formalism \cite{A}. Compared to the Dirac
approach this is basically a Lagrangean approach. Moreover it
does not require the classification of constraints which is
essential to the Dirac procedure. Details of this procedure in
the context of both unconstrained and constrained systems have
been provided elsewhere \cite{FJ, BW}. The essential idea is to obtain the
symplectic two form, the inverse of which yields the brackets of
the theory. For an unconstrained system this is reasonably
straightforward since the symplectic matrix is invertible. For
constrained systems a possible approach \cite{BW} is to find the zero
modes of the symplectic matrix which is no longer invertible. If
the zero modes do not produce any restriction on the dynamical
variables, it indicates the occurrence of a gauge symmetry. The
natural way to proceed then is to fix a gauge. In the present
model, however, explicit gauge fixing is not necessary. The
brackets of the theory are directly obtained by working with the
independent variables of the system. Since the symplectic analysis
is suited for first order systems, the Lagrangean (\ref{d4}) is
ideal for this purpose. Indeed, contrary to the usual Maxwell
Lagrangean, conversion to the first order form by introducing
auxiliary variables becomes redundant. 

To quickly recapitulate the basic tenets of the symplectic
approch,{\footnote {The general development is given for point
mechanics. The extension to field theory is self evident.}}
the geometric structure is induced by the closed
symplectic two form,
\begin{equation}
f^{(0)}=\frac{1}{2}f_{ij}^{(0)}d\omega_i^{(0)}d\omega_j^{(0)}
\label{s01}
\end{equation}
where,
\begin{equation}
f_{ij}^{(0)}=\frac{\partial a_j^{(0)}}{\partial\omega_i^{(0)}}
- \frac{\partial a_i^{(0)}}{\partial\omega_j^{(0)}}
\label{s02}
\end{equation}
and
$a^{(0)}(\omega^{(0)})=a_i^{(0)}(\omega^{(0)})d\omega_i^{(0)}$
is the canonical one form defined from the original Lagrangean,
\begin{equation}
L^{(0)} dt=a^{(0)}(\omega^{(0)})-V^{(0)}(\omega^{(0)}) dt
\label{s03}
\end{equation}
The superscript $0$ implies that the original Lagrangean is being
considered. In fact it is  indicative of the iterative nature of the
computations. Additional restrictions coming from the
constraints are imposed through
Lagrange mutipliers in which case one has to extend the
configuration space. The corresponding Lagrangean gets modified 
and accordingly the superscript also changes. The process
terminates once the symplectic matrix becomes invertible.

It is easy to see that the two form following from (\ref{d4}),
\begin{equation}
f^{ab}_{ij} = -{\epsilon_{ab}}{\epsilon_{ijk}}\partial_k\delta(x-y)
\label{s1}\end{equation}   
does not have an inverse. It is possible to generically denote
the  zero modes  by,
\begin{equation}
\nu_{ij}^{ab}= \partial_i\phi_j^{ab}
\label{s2}
\end{equation}
To check whether new conditions are generated by these modes,
one has to study the relation \cite{BW},
\begin{equation}
\int \nu\frac{\partial V}{\partial\omega} = 0
\label{s3}
\end{equation}
Using (\ref{s2}) it is seen that the condition (\ref{s3}) is
trivially satisfied. Hence there are no further restrictions and
the system is a gauge theory. It is also clear that (\ref{d4})
is expressed solely in terms of the transverse components of the
potentials so that the longitudinal components can be set equal
to zero. This is incorporated by modifying the Lagrangean as,
\begin{equation}
{\cal L} =\frac{1}{2}\epsilon^{jki}\partial_j{A_k}^{a}
\epsilon_{ab}\partial_{0}A^b_{i}- 
\frac{1}{4}F^{a,jk}F^a_{jk}+
 \dot\lambda^a\partial_iA^a_i
\label{s4}\end{equation}
Note that the Lagrange multiplier enforcing the transversality
has been introduced by means of a time derivative to simplify
the ensuing algebra. Physically this means that the
transversality condition is time independent. It should perhaps
be mentioned that the transversality condition enforced here is
not the same as implementing the Coulomb gauge in usual Maxwell
theory. In the latter case the longitudinal components are
forced to vanish by putting an external gauge condition. Here,
on the contrary, the duality invariant Lagrangean is already
expressed in terms of the transverse components and the
transversality condition separates out the independent
components. This will become more clear when we discuss the same
theory without introducing any Lagrange multiplier. The symplectic
variables are now $A_i, \lambda$. The first iterated symplectic
matrix is  given by,
\begin{equation}
f_{ij}^{ab} ={\pmatrix{\epsilon_{ab}\epsilon_{ijk}
\partial_k & -\delta_{ab}\partial_i
 \cr -\delta_{ab}\partial_j & 0}} \delta(x-y) 
\label{s5}
\end{equation}
The above matrix is nonsingular and its  inverse is given by,
\begin{equation}
f_{jl}^{ab(-1)} = {\pmatrix{\epsilon_{ab}\epsilon_{jlm}
{\frac{\partial_m}{\nabla^2}}
&
{-\delta_{ab}}{\frac{\partial_j}{\nabla^2}}  \cr
{-\delta_{ab}}{\frac{\partial_l}{\nabla^2}} & 0 }} \delta(x-y) 
\label{s6}
\end{equation}
The brackets among the basic variables are now easily read-off
from the first entry,
\begin{equation}
\{{A^{a}_{i}}(x) , {A^b}_j(y)\} ={\epsilon_{ab}}{\epsilon_{ijk}} 
{\frac{\partial_k}{\bf \nabla^2}} \delta(x-y)\label{s7}
\end{equation}
which reproduces the Dirac algebra (\ref{d11a}) and hence
(\ref{d11}) also.  The other
brackets (\ref{d12}) and (\ref{d13}) found in the Dirac approach
are nonexistant in the
symplectic formalism since canonical momenta are never introduced.

The above algebra can also be obtained  without
9enlarging the configuration space. This is achieved by directly
working with the independent transverse variables which are
isolated by choosing the following polarisation,
\begin{equation}
{A^{a}_{i,T}} = 
(\delta_{i\alpha} - \delta_{i3}{\frac{\partial_\alpha}{\partial_3}})
{a^{a}_{\alpha}};\,\,\, \alpha=1, 2
\label{s9}
\end{equation}
Note that $a_1$ and $a_2$ are the two independent variables.
Expressed in terms of these variables, the kinetic part of the 
Lagrangean (\ref{d4})
takes the form,
\begin{equation}
{\cal L}_{KE} = \frac{1}{2} \epsilon_{\alpha\beta} \epsilon_{ab}\Big(
\partial_3 a_\beta^a+
 2\frac{ \partial_\beta\partial_\sigma}{\partial_3}a_\sigma^a\Big) 
 \partial_0 a_\alpha^b
\label{s10}\end{equation}
Thus proceeding as before ,we get the symplectic  matrix as,
\begin{equation}
f^{ab}_{\alpha\beta} =
 {\pmatrix{0 & {\epsilon_{ab}}{\frac{\nabla^2}{\partial_3}}
 \cr -{\epsilon_{ab}}{\frac{\nabla^2}{\partial_3}} & 0}} \delta (x-y)
\label{s12}\end{equation}
Its inverse is given by,
\begin{equation}
f^{-1} =\pmatrix{0  &  \epsilon_{ab}
\frac{\partial_3}{\nabla^2} \cr  -\epsilon_{ab}
\frac{\partial_3}{\nabla^2} & 0 } \times \delta (x-y)
\label{s13}
\end{equation}
The only nontrivial bracket is given by the off-diagonal entry,
\begin{equation}\label{s14}
\{ a^{a}_{1}(x), a^{b}_{2}(y) \} ={ \epsilon_{ab}}{\frac{\partial_3}
{\nabla^2}}\delta (x-y)       
\end{equation}
It is easy to verify that this is the same algebra found earlier
in (\ref{d11}) or (\ref{s7}).

Before concluding this section we discuss the axial gauge
formulation. 
This gauge is imposed as,
\begin{equation}
{\cal L}^{(1)} = 
{ \frac{1} {2}}{\epsilon^{jki}\partial_j{A_k}^{a}}
{\epsilon_{ab}\partial_{0}{A^b_{i}}}- {\frac{1}{4}}{F^{a,jk}}{{F^a}_{jk}} +
{\dot \lambda^{(a)}}{A^{a}_3}
\label{s15}
\end{equation}
The above first iterated Lagrangean leads to a nonsingular
symplectic matrix. Inverting it the relevant brackets are
obtained. These correctly reproduce (\ref{n3}).

\bigskip

\section{\bf Conclusions}

The present paper has shown that there are several facets to the
constrained dynamics of duality symmetric Maxwell theory. An
intriguing feature was the appearance of a reducible set of 
second class constraints together with first class constraints,
whereas the normal Maxwell Lagrangean has
only the latter. Nevertheless by suitably accounting for the
reducibility using an enlargement prescription, Dirac brackets
among the physical variables were directly obtained.
In other words any explicit reference to either the first class
constraints or the gauge
fixing, which has been the standard approach \cite{G} using the
Coulomb gauge,
was avoided. This is also a distinct point of departure
from the conventional constrained analysis in the Maxwell theory
where the full set of constraints is required to calculate the
Dirac brackets. For the axial gauge, on the other hand, it was
necessary to take the complete set of constraints simultaneously.
In this case a close parallel with the usual Maxwell analysis was
established since similar canonical pairs, 
modified by proper internal indices, were obtained.

The Dirac anlysis was followed by a symplectic approach. Indeed
the first order nature of the duality symmetric Lagrangean
naturally lends itself to such an analysis. It is not surprising
therefore that the symplectic formalism yields results in a more neat and
compact fashion compared to the Dirac procedure. This may be
contrasted with the normal second order Maxwell Lagrangean where
the application of either the Dirac or symplectic formulations
is more a matter of taste. The results in the symplectic
approach were obtained both in the gauge independent and gauge
fixed versions and were shown to agree with the Dirac analysis.
As is characteristic of the symplectic formalism, no reference
was made to the nature of the constraints. In fact the natural
first order form of the Lagrangean simplified the problem
considerably  so that no constraints emerged from the analysis.
The gauge symmetry was shown to be a direct consequence of the
redundancy in the degrees of freedom.

A possible generalisation can be made for other duality
symmetric models. Since these are first order Lagrangeans, it is
obvious that the symplectic formalism would be ideal for
discussing the constrained structure for such models. It also
seems reasonable to presume that this constrained structure
would be quite distinctive from that of the respective parent
Lagrangeans. This opens up the possibility of obtaining new
features from the duality symmetric Lagrangeans and gives us an
added reason to study such systems which are quite independent
of confining ones attention to just probe the nature of duality symmetry.

\end{document}